\documentclass[aps,pra,reprint,twocolumn,superscriptaddress,,showpacs,showkeys,longbibliography]{revtex4-1}

\usepackage{amsmath,amssymb,amstext}
\usepackage[usenames,dvipsnames]{color}
\usepackage{graphicx}
\usepackage{bm,bbold,braket}
\usepackage{natbib}
\usepackage{txfonts, comment,stmaryrd}
\usepackage{dcolumn}

\usepackage[english]{babel}

\usepackage[colorlinks,bookmarks=false,citecolor=blue,linkcolor=red,urlcolor=blue]{hyperref}

\begin{document}

\title{Dark solitons generation and their instability dynamics in two dimensional condensates  }

\author{Gunjan Verma}
\affiliation{Indian Institute of Science Education and Research, Pune 411 008, India}
\author{Umakant D. Rapol}
\affiliation{Indian Institute of Science Education and Research, Pune 411 008, India}
\affiliation{Center for Energy Sciences, Indian Institute of Science Education and Research, 
Dr. Homi Bhabha Road, Pune 411 008, India}
\author{Rejish Nath} 
\affiliation{Indian Institute of Science Education and Research, Pune 411 008, India}


\begin{abstract}
We analyze numerically the formation and the subsequent dynamics of two-dimensional matter wave dark solitons in a Thomas-Fermi rubidium condensate using various techniques. An initially imprinted sharp phase gradient leads to the dynamical formation of a stationary soliton as well as very shallow grey solitons, whereas a smooth gradient only creates grey solitons. The depth and hence, the velocity of the soliton is provided by the spatial width of the phase gradient, and it also strongly influences the snake-instability dynamics of the two-dimensional solitons. The vortex dipoles stemming from the unstable soliton exhibit rich dynamics. Notably, the annihilation of a vortex dipole via a transient dark lump or a vortexonium state, the exchange of vortices between either a pair of vortex dipoles or a vortex dipole and a single vortex, and so on.  For sufficiently large width of the initial phase gradient, the solitons may decay directly into vortexoniums instead of vortex pairs, and also the decay rate is augmented. Latter, we discuss alternative techniques to generate dark solitons, which involve a Gaussian potential barrier and time-dependent interactions, both linear and periodic. The properties of the solitons can be controlled by tuning the amplitude or the width of the potential barrier. In the linear case, the number of solitons and their depths are determined by the quench time of the interactions. For the periodic modulation, a transient soliton lattice emerges with its periodicity depending on the modulation frequency, through a wave number selection governed by the local Bogoliubov spectrum. Interestingly, for sufficiently low barrier potential, both Faraday pattern and soliton lattice coexist. The snake instability dynamics of the soliton lattice is characteristically modified if the Faraday pattern is present.
\end{abstract}



\maketitle
\section{Introduction}
Dark solitons (DSs) are fundamental, localized nonlinear excitations appearing in various systems, \cite{ds_kiv_98,ds_88, ds_91, ds_90, ds_92,ds_93, ds_09} including Bose-Einstein condensates (BECs) \cite{panay_08, ds_bec_10,panay_15}, with defocusing non-linearity  and they are characterized by a localized dip in the background density. Various techniques have been employed to generate matter wave DSs, for instance, phase imprinting \cite{ds_99_bec,ds_00_bec,ds_prl_08,kai_jrsol_16}, density engineering \cite{ds_sci_01,ds_merg_08_1,ds_latt_03} and a combination of both, so called the quantum state engineering \cite{ds_gen_01, ds_gen_02,da_decay_01,ds_natphy_08}. In addition, the formation of DSs has been reported for experiments, in which a condensate is flowing past an obstacle realized via laser induced dipole potential \cite{ds_flow_07,ds_vr_ring_09}, as well as during the interference of condensates initially prepared in multi-well potentials \cite{ds_inter_08,ds_merg_3_07}. DSs may also appear spontaneously in condensates, if the bose gas is quenched across the BEC phase transition via Kibble-Zureck mechanism \cite{ds_zure_09,ds_zur_expt_13,ds_bec_PT_10}. Besides the experimental investigations, there has been numerous theoretical studies \cite{ds_th_bia_02,ds_the_kev_03,ds_engg_ring_16,ds_gen_theo_05, ds_gen_theo2_05,ds_gen_th_04} discussing novel ways to generate, and study solitons, including their unique properties.

At zero temperatures, the stability of a DS demands a quasi one-dimensional condensate, i.e., the transverse extension of the condensate is required to be less than or of the order of the healing length \cite{ds_1d_jack_98, ds_stab_trap_99,ds_bush_00,ds_carr_00}. Otherwise, the soliton becomes dynamically unstable against transverse excitations, so called the snake instability (SI). SI leads to the bending of the nodal stripe (plane) in a two-dimensional \cite{2D_ds_ins_95, ds_lumb_03} (three-dimensional) background, which eventually breaks up into vortices and sound excitations \cite{ds_SI_th_00}. It has been observed experimentally in both matter wave  \cite{ds_dec_exp_01} and optical DSs \cite{ds_opt_dec_96,ds_bre_prl_96}. At finite temperatures, thermodynamic instability may lead to the decay of dynamically stable DSs \cite{ds_ther_dec_02}. In the trapped case, even at zero temperature, DSs may exhibit dissipative losses due to the emission of sound waves and its repeated collisions with the latter \cite{ds_99_bec, ds_dec_par_03, ds_dec_04, ds_dec_pre_05}. In general, the sound waves mediated dissipation of DS may occur if it encounters an inhomogeneity in the background density \cite{ds_dec_jpb_03}. Non-conventional and exotic DS-like excitations are also predicted to exhibit in 2D condensates \cite{ds_lumb_03,ds_exo_04,ds_exo_05,jvor_16}.

A dark stripe undergoing SI decays into a chain of vortices with opposite topological charges or the so-called vortex dipoles (VDs), a problem which has been addressed in the context of both non-linear optics \cite{ds_opt_si_93, ds_dec_opd_93, ds_opt_dec_96,ds_bre_prl_96,ds_kiv_rep_00} and in atomic condensates \cite{ds_lumb_03,ds_dec_2d_04,ds_dec_bar_10,SI_tor_13}, including fermionic superfluids \cite{ds_fermi_13}. If the transverse extension of the condensate is not large enough, the formed VD recombines back to form the dark stripe and will undergo SI again. This process keeps repeating  in time \cite{ds_dec_2d_04, ds_vr_ring_09}. As the transverse size increases, more  vortices are formed along the nodal stripe. Complementary to SI dynamics, it has also been shown that the various vortex configurations can be seen as the bifurcating states from a DS stripe \cite{ds_bif_10}. The complex dynamics of vortices  \cite{sv_dyn_rosh97,v_dyn_bar_08,v_dyns_hall_13} including that of VDs \cite{vd_stable_tor_03,vd_ener_04,vd_dyn_stat_05,vd_dyn_vir_06,vd_dyn_07,vd_dyn_li_08,v_dyn_bar_08, vd_dyna_sci_10, vdi_dyn_11, vd_dyn_kuo_11,vd_dyn_midd_11,vd_dyn_prl_10,vd_shed_15,vd_dyn_good_16} in an inhomogeneous interacting condensate is a subject of its own merit. The motion of a vortex is strongly affected by the interaction with other vortices and the geometrical confinement. For instance, two vortices with same topological charge precess around each other, whereas with opposite charges move together with a linear velocity along the direction of flow between them. A VD may annihilate, \cite{vd_dyn_07,vd_dyn_prl_10,vd_anni_ango_13,vd_dec_tub_14,vd_dec_stag_15, vd_dec_jhep_15} upon collision with other vortices or by emitting sound waves \cite{v_anni_so_13}, a process that might play a big role in realizing the Onsager vortex states \cite{ons_vd_dec_14, vort_gro_16}.

\begin{figure}[hbt]
\centering
\includegraphics[width= .9\columnwidth]{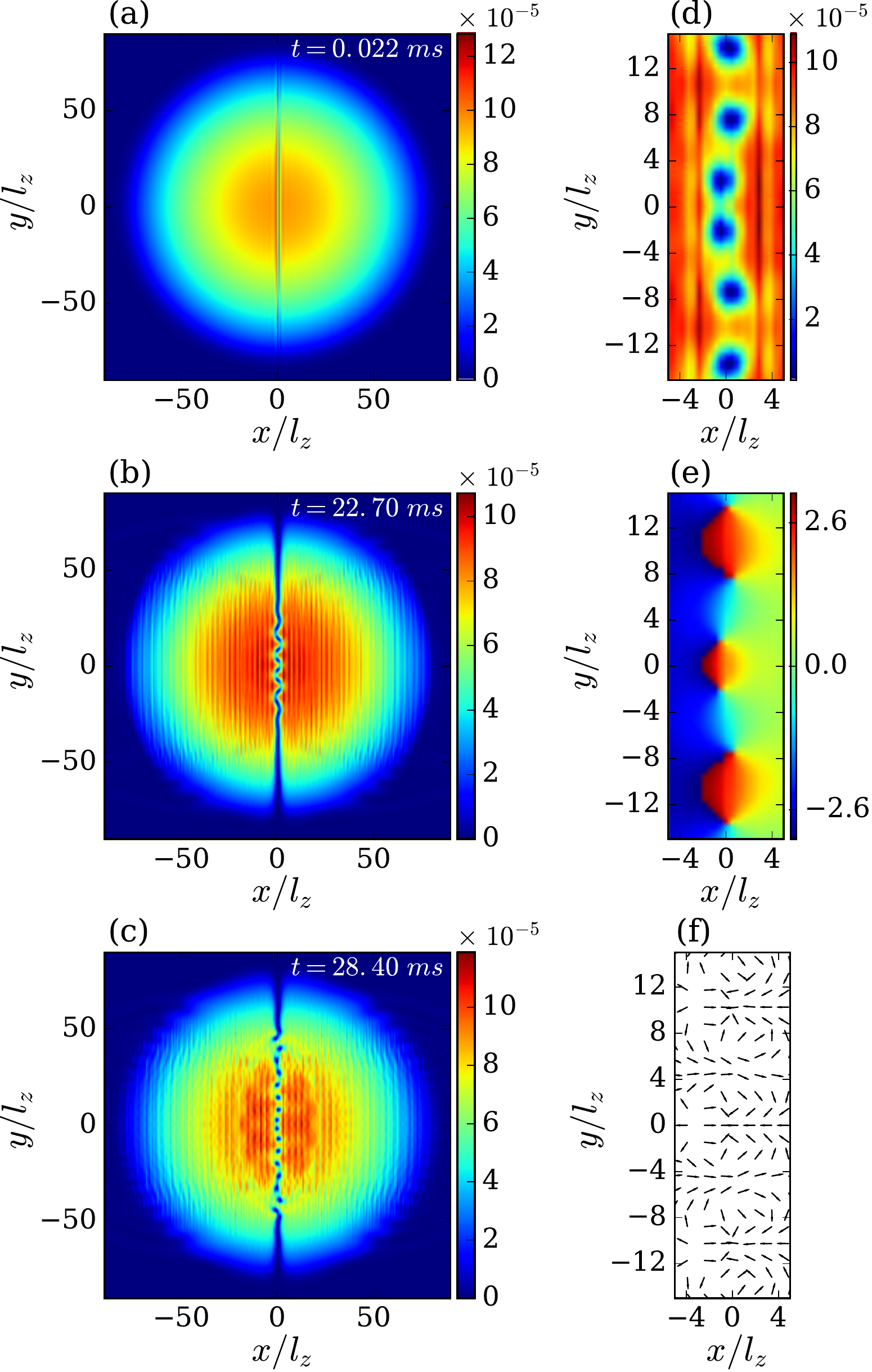}
\caption{\small{(Color online) The time evolution of the condensate density $n(x,y,t)=l_z^2|\psi(x,y,t)|^2$ of a rubidium BEC with $N=100$k, $a=109 a_0$, $\omega_x=\omega_y=2\pi\times 10$ Hz and $\omega_z=2\pi\times 700$ Hz, with an initial phase: $\theta(x)=\pi/2$ for $x<0$ and  $\theta(x)=-\pi/2$ for $x\geq 0$. (a) The condensate density at $t=22\ \mu$s shows the formation a nodal stripe which eventually transforms into a stationary DS. (b) The SI first develops at the centre of the condensate. (c)  A transient stripe of VDs emerge, and in (d) we show the VDs at the centre of the condensate. The phase and velocity fields corresponding to (d) are shown in (e) and (f) respectively.}}
\label{fig:d1} 
\end{figure}

In the first part of the paper, we numerically study the long-time dynamics of a 2D matter wave DS in a rubidium Thomas-Fermi (TF) condensate, subjected to SI. In particular, we look at the long-time dynamics of the VDs created. The DS is developed by imprinting a phase gradient at the start of the time evolution. The soliton properties, as well as the nature of SI and post instability dynamics, depend crucially on the spatial width of the phase gradient. A sharp phase variation at the centre of the condensate creates a stationary DS along with very shallow grey solitons, and the former eventually undergoes SI, results in a transient VD stripe. The stripe then breaks up into an interacting gas of VDs and lone vortices. The dynamics of vortices provides us very interesting scenarios: the annihilation of a VD via an intermediate dark lump or vortexonium state, the collisional dynamics resulting in the exchange of vortices between either two VDs or a VD and a single vortex, etc. Interestingly, for the sufficiently large width of the initial phase gradient, the solitons may decay directly into vortexoniums instead of vortex pairs. In the final part, we discuss  alternative techniques to generate  DSs in atomic condensates and the emerging instability dynamics. Especially, we focus on methods involving a Gaussian potential barrier and time-dependent interactions, both periodic and a linear quench. The properties of the solitons can be controlled by tuning the amplitude or width of the potential barrier. In the linear case, the number of solitons and their depths depend on the quench time of the interactions. Interestingly, for the periodic case, a transient DS lattice emerges and depending on the strength of the potential barrier it may coexist with Faraday patterns (FPs). The periodicity of the soliton lattice can be tuned easily by the modulation frequency, through a wave number selection governed by the local Bogoliubov excitations. The characteristics of the lattice instability dynamics depend crucially on whether there is a FP or not.

The paper is structured as follows. In Sec. \ref{si} we analyze numerically the SI dynamics of DSs in 2D TF condensates, in particular, the dependence of the spatial extension of the phase gradient. In Sec. \ref{vds}, we discuss the post-SI dynamics including that of the emanated VDs and vortexoniums. We discuss alternative techniques for generating DSs and the corresponding instability dynamics in Sec. \ref{atec}, including how to control the soliton properties by tuning the system parameters. Finally, we conclude in Sec. \ref{con}.

\section{Snake instability}
\label{si}
DSs can appear as defects in the background  density if we drive the initial state of the condensate into a dynamically unstable configuration. A simple way to achieve this is by imprinting a phase gradient in the condensate wave function, and  in Sec. \ref{atec} we discuss alternative techniques to achieve it. In this section, we focus mainly on the real-time dynamics of a 2D TF condensate with a phase gradient imprinted on it. We consider a  BEC of $N$ atoms confined in a harmonic trap: $V_t({\bf r})=m(\omega_x^2x^2+\omega_y^2y^2+\omega_z^2z^2)/2$ where $\omega_{\alpha}$ is the trap frequency along the $\alpha$ axis. To simplify, we take $\omega_z\gg \omega_{x,y}$ such that the dynamics of the condensate along the $z$-axis is frozen and the system can be effectively described by a two-dimensional (2D) Gross-Pitaevski equation (GPE):
\begin{equation}
i\hbar\frac{\partial \psi(x,y, t)}{\partial t}=\left[\frac{-\hbar^2\nabla^2}{2m}+V_t(x,y)+g_{2D}|\psi(x,y, t)|^2\right]\psi(x,y, t),
\label{gp2de}
\end{equation}
where $\psi(x,y, t)$ describes the condensate wave function in the $xy$ plane, $V_t(x,y)$ is the radial part of the harmonic confinement and $g_{2D}=g/\left(\sqrt{2\pi}l_z\right)$ is the effective coupling constant in 2D with $g=4\pi\hbar^2aN/m$. The validity of Eq. \ref{gp2de} demands $l_z < \xi$ where $l_z=\sqrt{\hbar/m\omega_z}$ and $\xi$ is the healing length \cite{lowd_con_01}. We solve Eq. \ref{gp2de} in real time starting with the initial solution $\psi(x,y,t=0)=\psi_0(x,y)\exp[i\theta(x)]$, where $\psi_0(x,y)$ is the ground state solution of Eq. \ref{gp2de} obtained via imaginary time evolution and $\theta(x)$ is the $x-$dependent initial phase. By introducing a phase gradient along the $x$-axis, the dark solitons are generated with nodal lines parallel to the $y$-axis \cite{ds_99_bec,ds_phase_01}.  First we take $\theta(x)=\pi/2$ for $x<0$ and  $\theta(x)=-\pi/2$ for $x\geq 0$, which introduces a sharp gradient at $x=0$ with a phase difference of $\pi$. The resulting dynamics of the condensate is shown in Fig. \ref{fig:d1}(a)-(f). In the initial stage, a nodal stripe is developed [see, Fig. \ref{fig:d1}(a)]  which eventually turns into a stationary DS at the centre of the BEC, accompanied by sound waves and very shallow gray solitons. The former then undergoes SI [see, Fig. \ref{fig:d1}(b)], which leads to the formation of a transient stripe of VDs [see, Fig. \ref{fig:d1}(c)] and additional sound excitations \cite{ds_SI_th_00, ds_dec_exp_01}. The SI has been attributed to the appearance of imaginary \cite{ds_stab_trap_99} or complex modes  \cite{ds_SI_th_00} in the Bogoliubov spectrum of a DS. Note that, the emanated sound waves during the phase imprinting feed up the unstable modes and enhance the SI. In addition, we observe the formation of a large number of gray secondary solitons.  There are other ways of creating shallow solitons in condensates, for instance, by generating shock waves, which shed DSs or by perturbing a stationary DS \cite{pert_ds_91}. Typically, very shallow DSs  disintegrate quickly after hitting the boundary.

\begin{figure}[hbt]
\centering
\includegraphics[width= 1.\columnwidth]{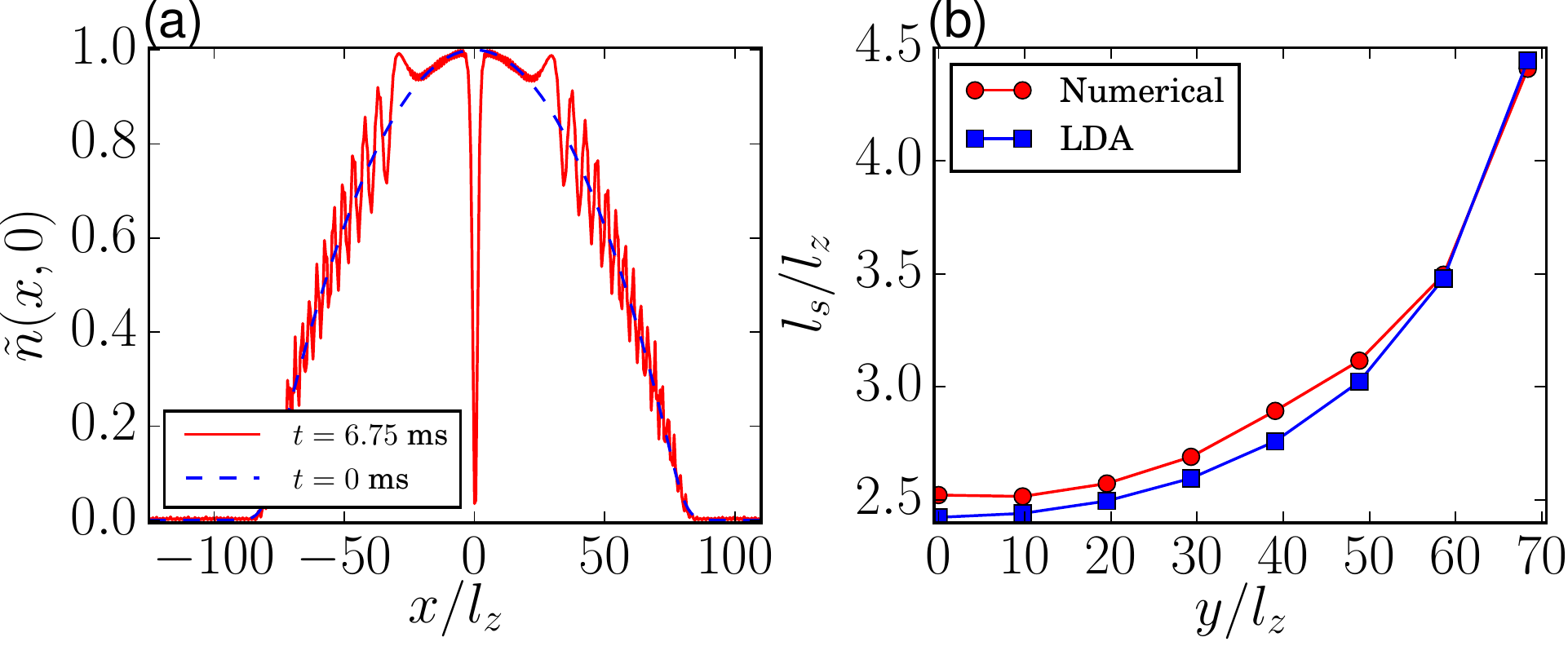}
\caption{\small{(Color online) (a) The TF density profile at $t=0$ (dashed line) and the emerged solitons at $t=6.75$ {\rm ms} (solid line). All plots are renormalized with its corresponding maximum density. (b) The numerical results (filled circles) for the width $l_s(y)$ of the stationary DS as a function of $y$ is compared with that of LDA with a TF profile (filled squares). The BEC setup is same as that of the Fig. \ref{fig:d1}.}}
\label{fig:d2} 
\end{figure}

Since the size of the condensate is much larger than the width ($l_s$) of the DS, we can describe the stationary soliton using the local density approximation (LDA)\cite{ds_lda_04,ds_lda_06}. The width $l_s$ of a stationary DS in a homogeneous background is provided by the healing length $\xi$ and is $\propto 1/\sqrt{n_h}$, where $n_h$ is the uniform density of the condensate. Using the LDA, we can approximate the $y$-dependent width of DS as $l_s(y)=\hbar/\sqrt{mg_{2D}n_0(y)}$ \cite{hl_NJP_15} calculated at $x=0$, where $n_0(y)=|\psi_0(x=0,y)|^2$. If we assume a Thomas-Fermi (TF) profile, the condensate density becomes $n_0(x,y)=\mu/g_{2D}\left[1-x^2/R_x^2-y^2/R_y^2\right]$ for $\{x\leq R_x \  \& \  y\leq R_y\}$ otherwise $n_0(x,y)=0$ with $R_{x,y}^2=2\mu/\left(m\omega_{x,y}^2\right)$. As we move away from the centre towards the edge of the condensate $l_s$ increases as a function of $y$ since the density decreases. In Fig. \ref{fig:d2}, we compare the numerical results for $l_s$ with that of LDA using TF profile. They are found to be in excellent agreement. For the numerical result, the $l_s$ is calculated at an instant of time at which the nodal stripe has completely developed into a stationary soliton, and after which the width of the DS hardly changes before undergoing SI. At this instant of time, the envelope of the condensate is almost identical to that of initial TF profile. To extract the soliton width, we crop density profile around the DS, and then re-normalize with the maximum density, which gives us a scenario identical to a stationary soliton in a homogeneous back ground. Then, the DS width is defined as the distance over which the density grows from 0 to 0.58 times the maximum.
\begin{figure}[hbt]
\centering
\includegraphics[width= .9\columnwidth]{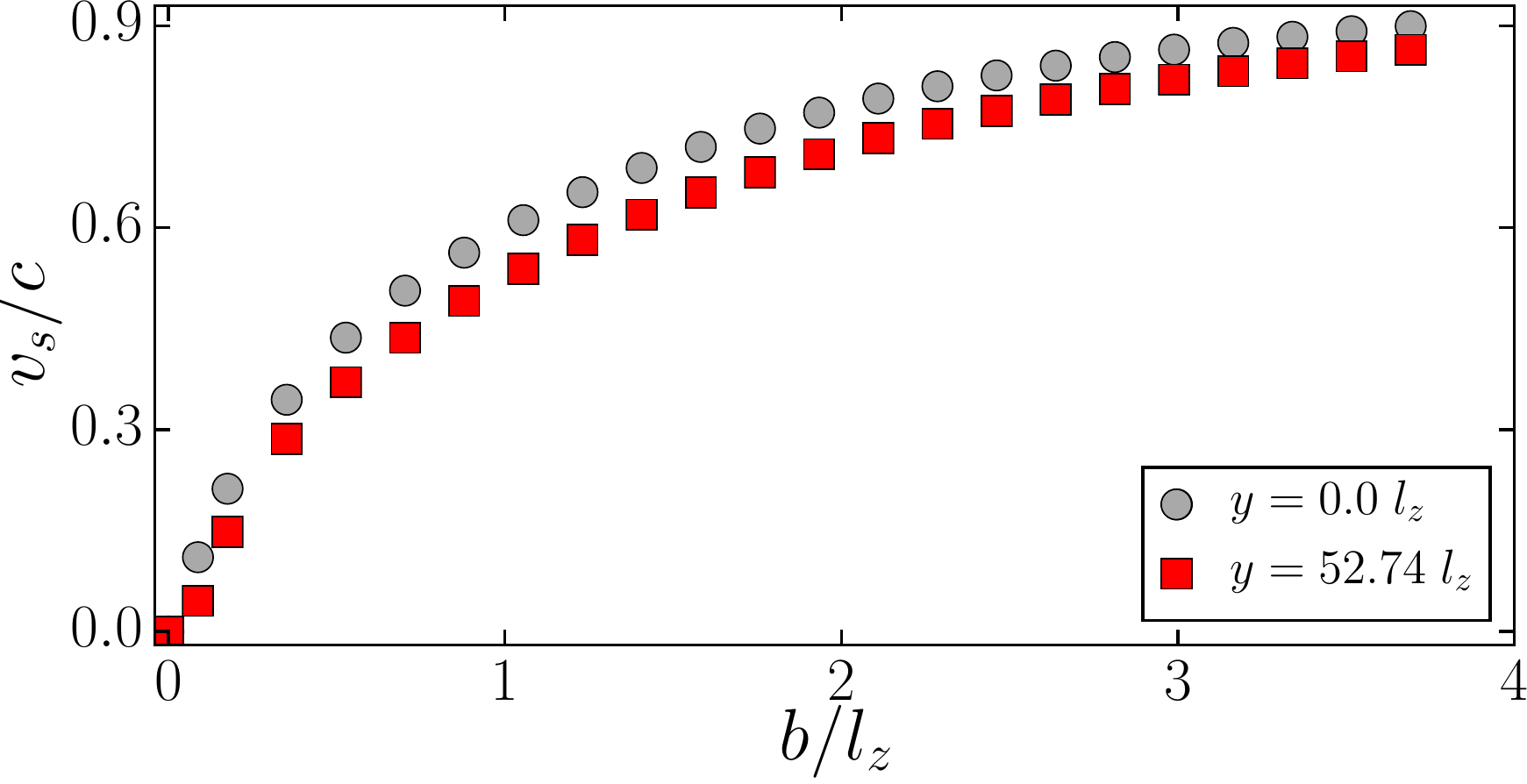}
\caption{\small{(Color online) The subsonic local velocity $v_s$ along the $x$-axis of the dark soliton as a function of $b$ at different locations along the $y$-axis. For both curves, we calculated the velocity at $x=1.5l_z$. At large values of $b$, $v_s$ saturates to the speed of sound $c$, and also the centre of the soliton moves faster compared to the edges. The setup is same as that in Fig. \ref{fig:d1}.}}
\label{fig:v} 
\end{figure}
\begin{figure*}[hbt]
\centering
\includegraphics[width= .9\textwidth]{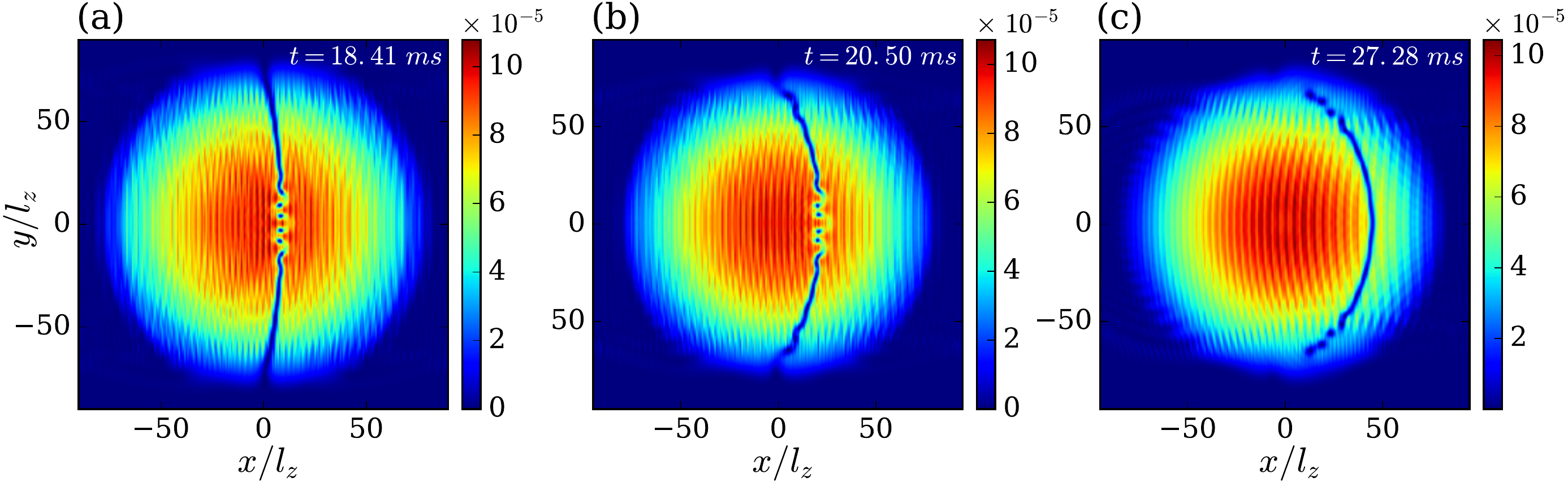}
\caption{\small{(Color online) The SI dynamics for different $b$ is depicted through the condensate density $n(x,y,t)$. (a) The SI develops first at the centre for $b=0.05 l_z$. (b) The SI develops both at the centre and the edges simultaneously for intermediate value of $b=0.2 l_z$. (c) For sufficiently large values of  $b=0.75 l_z$ the instability occurs first at the edges. In the later stage the whole DS breaks up into vortex dipoles. Note that, larger the $b$ shallower the soliton is, hence faster it propagates. The setup is same as that in Fig. \ref{fig:d1}.}}
\label{fig:sib} 
\end{figure*}

 Further, we have noticed that the SI of DS develops first at the centre of the condensate [see Fig.\ref{fig:d1}(b)] and later it spreads across the entire soliton. We explain this behaviour using LDA as follows. The SI in a homogeneous background is attributed to the imaginary low-momentum transverse excitations \cite{ds_stab_trap_99} as stated above. They appear for momenta $0\leq q\xi \leq 1$ with a maximum value of $\mu_h/4$ at $q\xi=1/\sqrt{2}$, where $q$ is the quasi-momentum and $\mu_h=g_{2D}n_h$ is the chemical potential of the homogeneous density. The maximum imaginary mode provides us the growth rate and hence determines the time scale at which the SI develops, which is $\tau=4\hbar/\mu_h$. Using LDA for the inhomogeneous case, we get $\tau(y)=4\hbar/[g_{2D}n_0(y)]$, which indicates that the instability grows faster at the centre compared to that in the edge. 

{\em Smooth phase gradient}: at this point, we replace the sharp phase gradient with a smooth one, $\theta(x)=(\pi/2)\tanh(x/b)$, which varies over a width $b$ \cite{ds_th_bia_02}. Interestingly, the  characteristics, SI and post-SI dynamics of the DS depend crucially on the parameter $b$. Note that, $b=0$ corresponds to the sharp gradient scenario discussed above and in that case, a stationary DS is created at the centre of the condensate. When $b$ is nonzero, the phase becomes continuous at $x=0$ and as expected, instead of a stationary soliton, a primary DS with a nonzero subsonic $y$ dependent velocity $v_s(y)\approx c[1-n_d/n_0(y)]^{1/2}$ along the $x$-axis is created, where $c=\sqrt{gn_0^{MAX}/m}$ is the speed of sound calculated at the maximum density $n_0^{MAX}$ and $n_d$ is the density depth of the DS.  Since $n_d$ decreases with increase in $b$, the velocity $v_s$ increases, as shown in Fig. \ref{fig:v}. At large values of $b$, the velocity $v_s$ saturates to $c$ \cite{ds_99_bec}. The density dependence of the soliton velocity results in the bending of the nodal stripe during time evolution [see Fig. \ref{fig:sib}], which is unrelated to the SI dynamics.  

\begin{figure}[hbt]
\centering
\includegraphics[width= 1.\columnwidth]{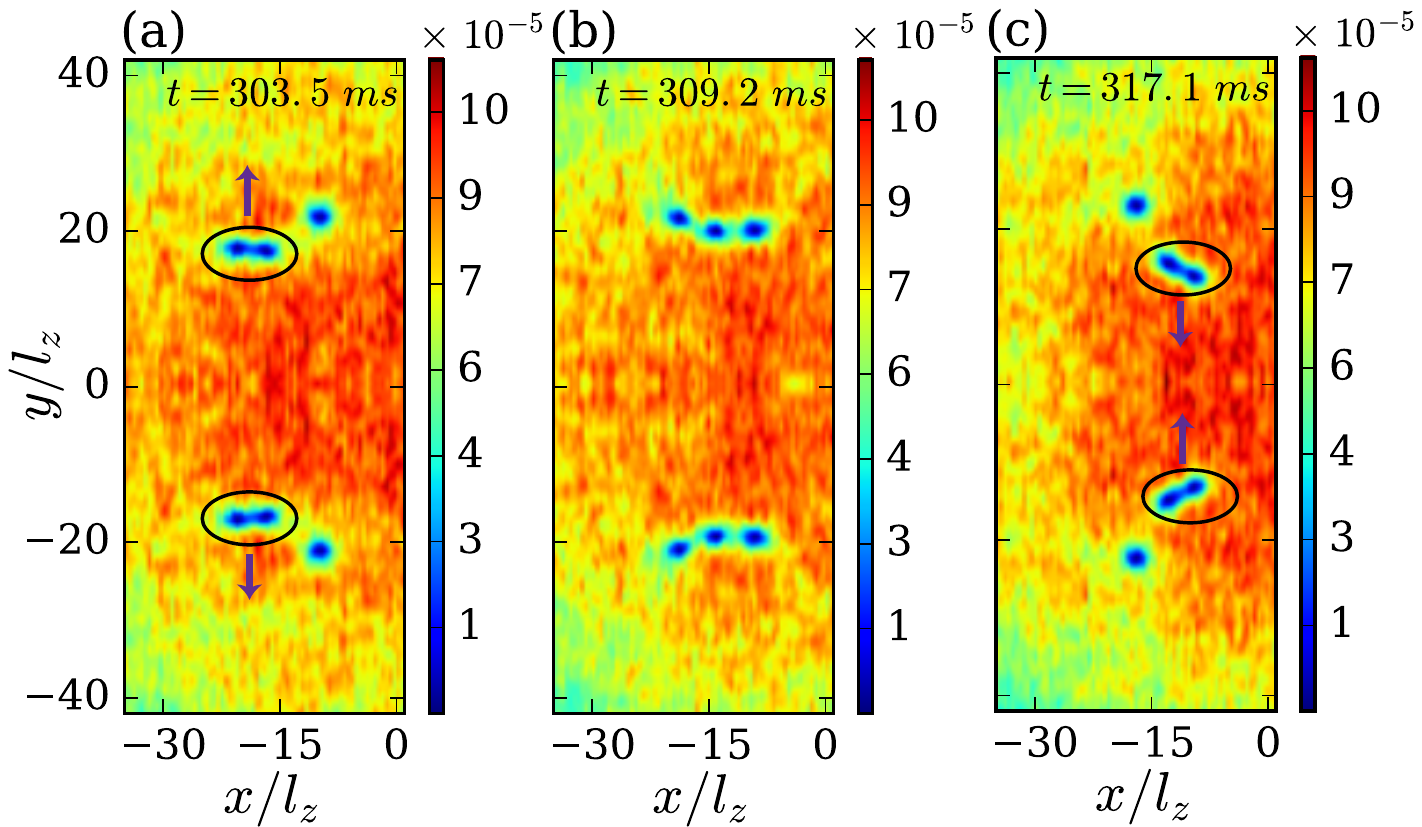}
\caption{\small{(Color online) (a) Two lone vortices are approached by two vortex dipoles and in (b) when they reach close to each other. In (c), an exchange of vortices has been taken place and a new pair of vortex dipoles are created. The setup is same as that in Fig. \ref{fig:d1} with an initially imprinted sharp phase gradient. The initial and final VDs are indicated by ellipses and direction of propagation of VDs are shown by arrows.}}
\label{fig:ve1} 
\end{figure}

The SI dynamics is also strongly modified by the parameter $b$. For small  $b$, the instability process is analogous to that of the stationary case discussed above, in a way that the instability grows first at the centre of the soliton stripe, see Fig. \ref{fig:sib}(a), then progress to the rest of the stripe. In contrary, as $b$ increases,  the instability appears simultaneously at the centre as well as at the edges, leaving an intermediate fragment of DS unaffected. Hence, by the time SI develops in the intermediate region, the vortex dipoles are already created at the centre and the edges, see Fig. \ref{fig:sib}(b). If $b$ is further augmented, the instability occurs first at the edges as shown in Fig. \ref{fig:sib}(c), then propagate towards the centre. Later, the entire soliton breaks up into VDs. Note that for  sufficiently large values of $b$, very shallow solitons are formed with $v_s\sim c$ which dissipate before the SI develops. 


\section{Vortex dynamics}
\label{vds}
The extended time evolution of the phase imprinted condensate reveals very rich and interesting scenarios in the vortex dynamics, especially of VDs. A stationary DS breaks up into an unstable configuration of a stripe of VDs, which eventually disperses into a gas of interacting vortices and VDs. A fraction of vortices breaks up from the centre of the VD stripe disperse towards the trap boundary (low-density region). Then, they precess along the trap boundary with either clockwise or anti-clockwise rotation depending on their vorticity, but a complete orbital motion \cite{vd_dyn_prl_10} is prevented by the presence of other vortices. In the course of time, we observe the following frequent events:
\begin{enumerate}
\item vortex pair annihilation, where two vortices of opposite charge approach each other, then coalesce into a neutral dark lump \cite{ds_lumb_03,ds_2d_vor_col_15} (the same is termed as vortexonium \cite{vort_gro_16}) which subsequently decay into sound excitations. Vortexonium is a spatially localized state, characterized by a phase step identical to that of a gray soliton, and it may revert to vortex dipole by hitting the boundary. VD annihilation has been reported in a BEC experiment \cite{vd_dyn_prl_10}, including in the context of superfluid turbulence \cite{vd_dec_tub_14}. The vortexonium may annihilate or decays to a smaller one, in the course of collision with a vortex or a VD. Those vortexoniums which form at the trap boundary may propagate towards the centre of the condensate and decays into density excitations.  
\begin{figure}[hbt]
\centering
\includegraphics[width= 1.\columnwidth]{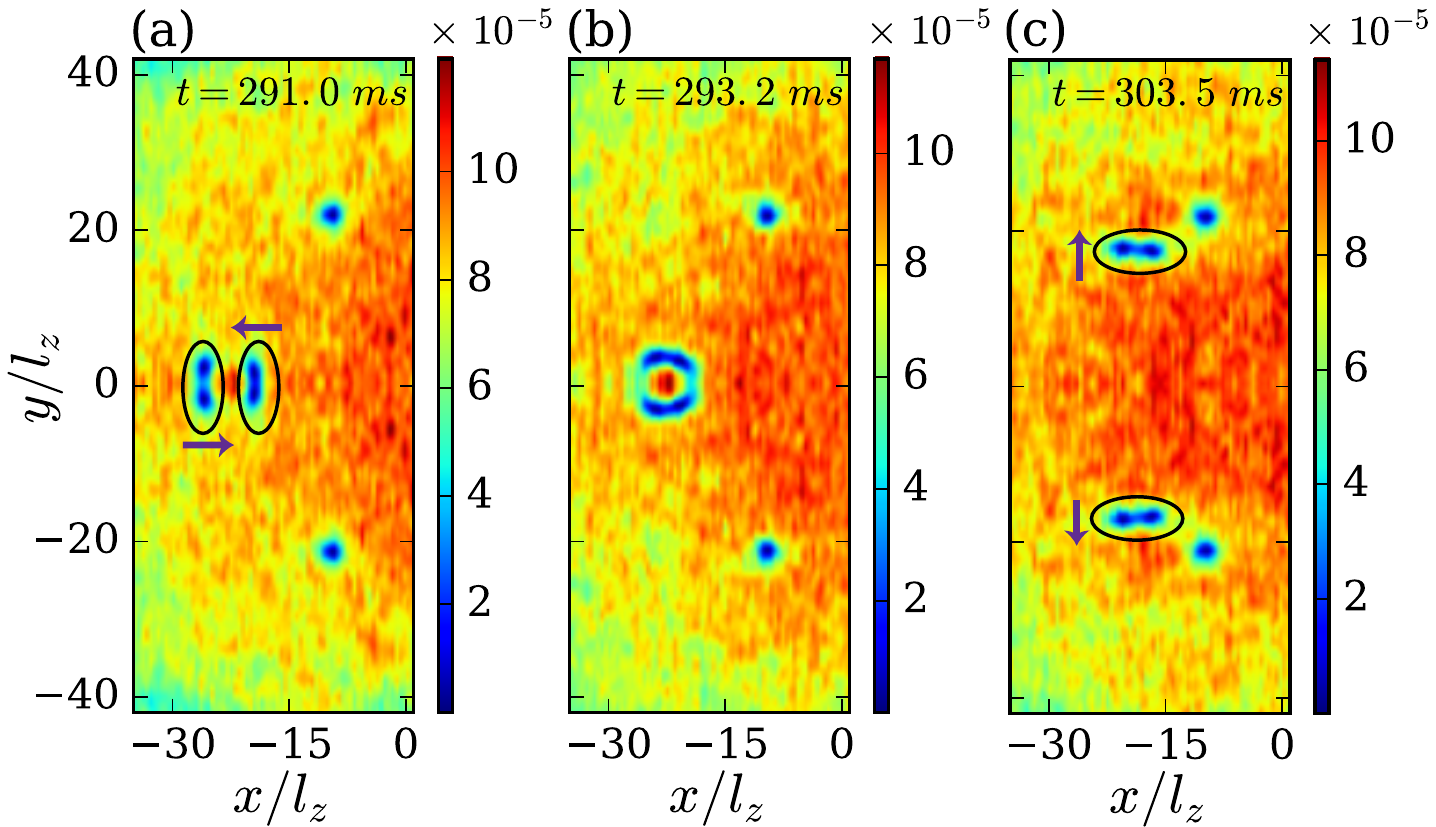}
\caption{\small{(Color online) (a) Collision of two counter propagating VDs. (b) Upon collision they exchange vortices and form a new pair of VDs. The new pairs propagate opposite to each other in a direction perpendicular to the axis of collision, as shown in (c). The initial and final VDs are indicated by ellipses and direction of propagation of VDs are shown by arrows.}}
\label{fig:ve2} 
\end{figure}
\item Among vortexoniums we found two scattering events, in one case two of them collide with each other, then emerge out intact and move away \cite{ds_lumb_03}. In the second case, after the collision one decays while the other breaks up into a VD. It is difficult to pinpoint on what condition each of these processes occurs independently and may depend on the kinetic energy of the vertexoniums at the instance of collision.
\item A VD, after colliding with a lone vortex, form a new VD and a new single vortex \cite{park_the_04,ds_2d_vor_col_15}. Two such process happening simultaneously are shown in Fig. \ref{fig:ve1}. The nature of dynamics, either flyby or exchange processes, depends on the relative velocity and the impact distance between theVD and the single vortex \cite{ds_2d_vor_col_15}. While approaching the single vortex, the VD slows down (also the energy of VD increases) hence, undergoing structural modifications, in particular, the vortex-anti vortex separation increases. Finally the antivortex of the original VD merges with the single vortex forming a new VD, accompanied by radiative losses as sound excitations. The new VD propagates in a direction opposite to that of the initial one resulting in a {\em back scattering}.
\item A pair of VDs  upon head to head collision, break up and exchange their partners, leading to the formation of a  new pair of VDs, see Fig. \ref{fig:ve2}. The new ones move perpendicular to the initial direction of propagation, as well as opposite to each other.
\item Two VDs approaching each other would either convert into two vortexoniums or annihilate one among them upon collision.
\end{enumerate}

\begin{figure}[hbt]
\centering
\includegraphics[width= .9\columnwidth]{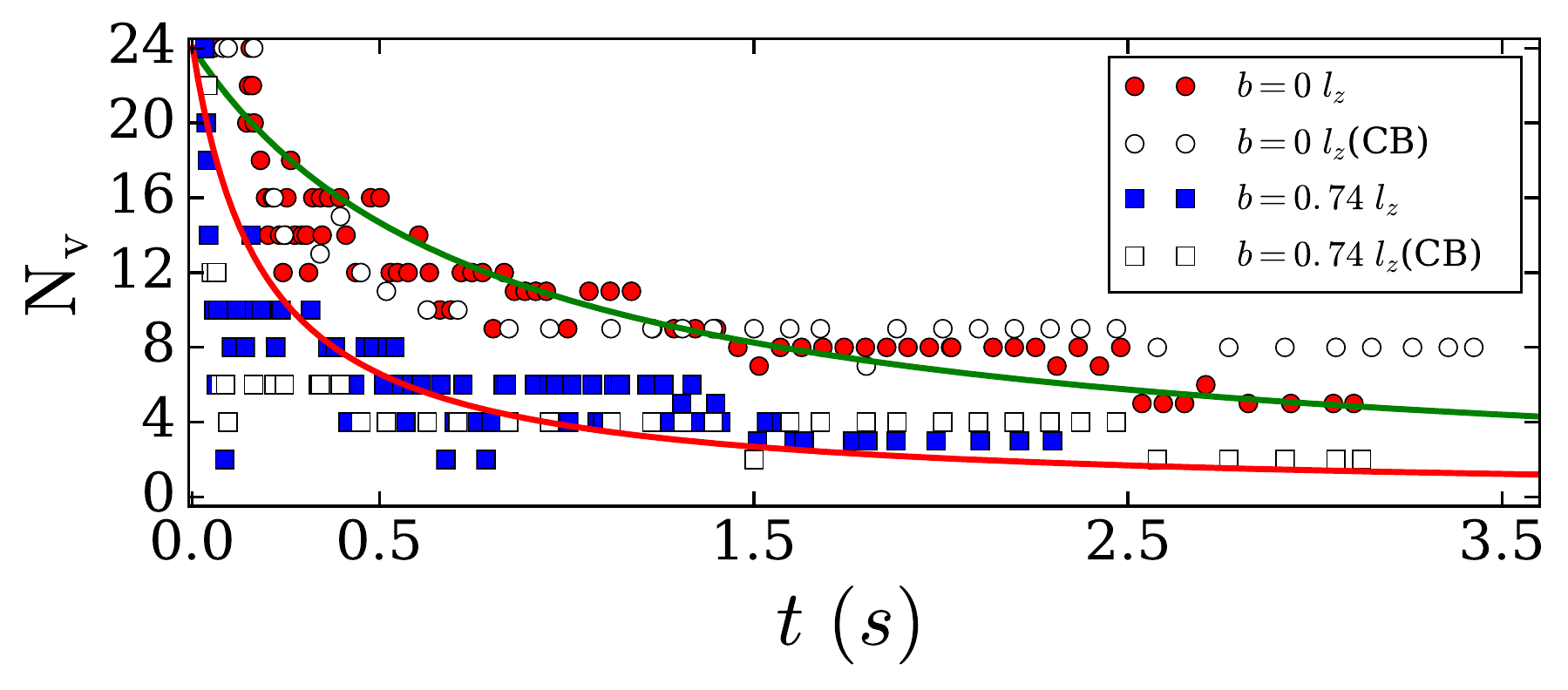}
\caption{\small{(Color online) The number of vortices as a function of time for two different $b$. Larger the value of $b$, faster the decay of vortices. The filled circles and squares are obtained by starting from a TF solution with an imprinted phase, whereas the empty circles and squares are for the cases in which time evolution starts with the dark solitons in a clean background (CB). The solid lines are calculated using the rate equation given in Eq. \ref{re} with $\Gamma(0)=0.053{\rm s}^{-1}$ and $\Gamma(0.74 l_z)=0.22{\rm s}^{-1}$. The oscillations in the vortex decay curve is due to the transitions between vortexoniums to VDs and vice versa. The setup is same as that in Fig. \ref{fig:d1}. The origin of the time is taken at the point at which the DS completely breaks up into VDs, which is of the order of milli seconds.}}
\label{fig:vdb} 
\end{figure}

The parameter $b$ indirectly influences the vortex dynamics. In particular, for sufficiently large b, the solitons first break from the edges into VDs, as mentioned above, then they quickly coalesce into vortexoniums. Finally, two adjacent vortexoniums collide, and one of them gets annihilated. This process repeats, resulting in the decline of the number of vortices as time progresses. We observed that as $b$ increases the annihilation process is more frequent causing a faster decrement in vortex number, see Fig. \ref{fig:vdb}. The oscillations in the vortex decay curve is due to the transitions between vortexoniums and VDs. The effective decay in the number of vortices, $N_v$, is well captured by the phenomenological rate equation \cite{vd_dec_tub_14},
\begin{equation}
\frac{dN_v}{dt}=-\Gamma (b) N_v^2,
\label{re}
\end{equation}
where the decay coefficient $\Gamma (b)$ depends on $b$. The results according to the rate equation is shown as solid lines in Fig. \ref{fig:vdb}. Another interesting feature we noticed is that, at sufficiently large $b$, DS travels faster, and breaks directly into vortexoniums at the boundary. This indicates that VDs are not necessarily the direct consequence of SI in 2D, depending on the nature of DS and its background density, it may break directly into vortexoniums. 

 Note that sound excitations are generated during the nucleation of  DSs, after the phase imprinting. To understand their role in the dynamics discussed above, we numerically look at the real time evolution starting from the stationary DS solution of the GPE. The SI is then triggered by a very small initial random noise, which is negligibly small compared to the sound excitations emerge from the phase imprinting. In contrary to the dynamics of the phase-imprinted condensates, here we observe two effects. (i) The SI has slow down, since in the phase-imprinted case the presence of sound excitations enhanced the SI, and (ii) the vortex-anti vortex annihilation process is faster, which is evident from the initial stages of the dynamics, as shown in Fig. \ref{fig:vdb}, which also supports the results that the presence of density perturbations may suppress the annihilation processes \cite{vd_anni_ango_13}. 


\section{Alternative techniques for soliton generation}
\label{atec}
In this section, we discuss alternative techniques for the generation of DSs in BECs and the corresponding dynamics in a 2D condensate. The methods involve the use of  an external potential barrier and time-dependent interactions, $g_{2D}(t)$. A potential barrier can be induced by an optical dipole potential or a far-detuned laser field. Generation of 1D solitons due to the sudden removal of the potential barrier is theoretically studied in \cite{ds_gen_02,ds_gen_Gobs_07}, and is also experimentally demonstrated \cite{ds_merg_08_1,ds_flow_07,ds_vr_ring_09}. It has shown that the creation of DSs require the system to be in a non-linear regime where the interaction energy dominates the kinetic part \cite{ds_coll_form_98, ds_vr_ring_09}. The characteristics of the solitons also depend crucially on the initial phase difference between the disconnected condensates, for instance, a phase difference of $\pi$ may lead to the formation of a stationary soliton \cite{ds_coll_form_98} accompanied by very shallow solitons, similar to that discussed in Sec. \ref{si}. The number of solitons can also be controlled by fine tuning the initial separation between the condensates. 

\subsection{Gaussian barrier and time dependent interaction}
We consider the same setup of BEC as in Sec. \ref{si} with an additional Gaussian $x$-dependent potential barrier : $V_G(x)=V_0\exp(-x^2/2d^2)$  of width $d$ and height $V_0$, at the centre of the trap. Hence, effectively we have a double well potential along the $x$-axis. Now, introducing the time-dependent interaction leads to the dynamical formation of solitons. In other words, the nodal stripe which is not a DS created by the potential barrier acts as a source of DSs in the presence of $g_{2D}(t)$.  Below, we consider both  linear  and periodic variation of $g$ in time \cite{gmod_10}. Note that, any non-zero $V_0$ breaks the radial symmetry of the condensate, and for sufficiently large $V_0$ and $d$, the condensate splits symmetrically into two disconnected BECs along the $x$-axis.

\subsubsection{Linear variation of $g$ in time}
\begin{figure}[hbt]
\centering
\includegraphics[width= 1.\columnwidth]{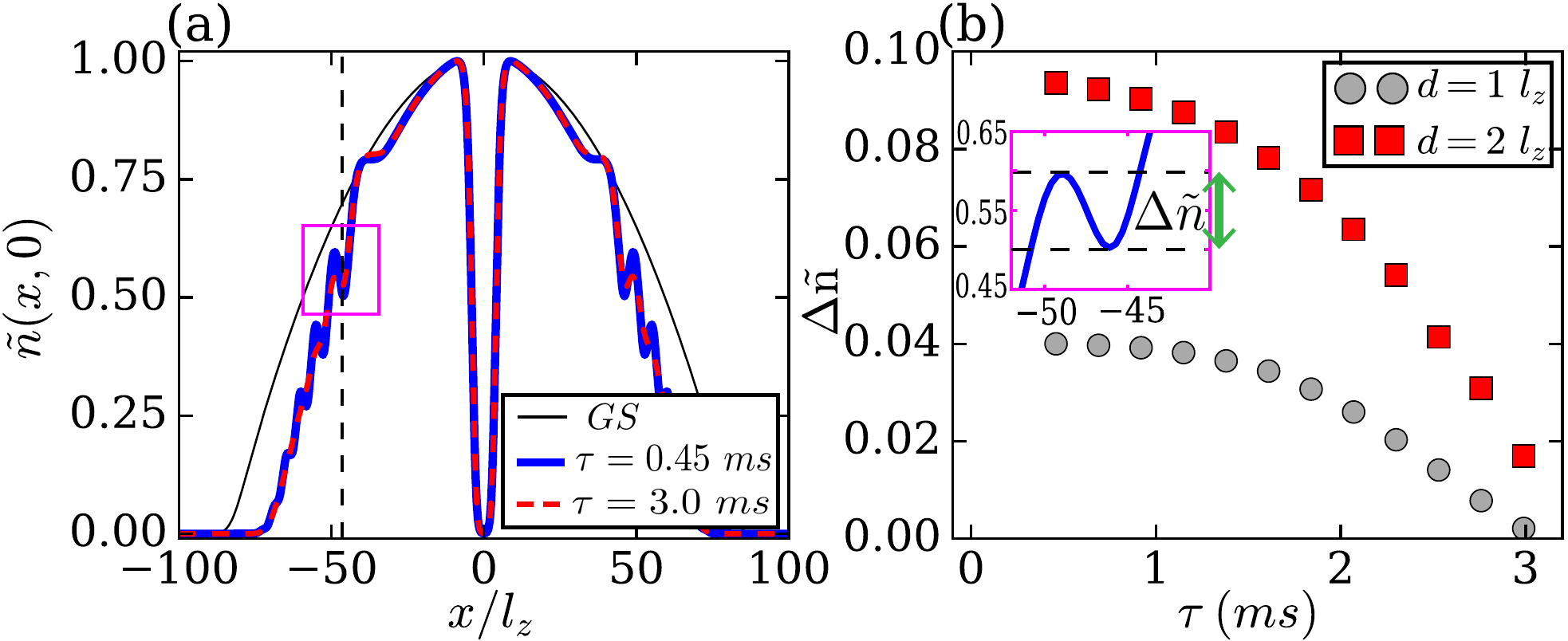}
\caption{\small{(Color online) (a) The density profiles for the ground state (GS)  (thin solid
line) and the emerged solitons for different ramp times, $\tau=1/\alpha= 0.45$ ms (thick solid line) and $\tau= 3$ ms (dashed line) are shown. All plots are renormalized with its corresponding maximum density and the snapshots are taken at different instants of time for different $\tau$. (b) Relative soliton depth (filled circles) for the width $d = 1 l_z$ and $d=2 l_z$ (filled squares) as a function $\tau$ is plotted. The soliton depths are calculated at the same location [$x=46.5 l_z$, shown with a box in (a)] in the condensate, shown by dashed vertical line in (a). The definition of $\Delta\tilde n$ is schematically shown in the inset of (b), which corresponds to the box in (a). The BEC setup is same as that of the Fig. \ref{fig:d1}, and for both (a) and (b) we take $\delta g=-g_{2D}/2$.}}
\label{fig:nd} 
\end{figure}

Here, we consider the linear variation of scattering length $a$ in time $t$, such that
\begin{equation}
g_{2D}(t)=\begin{cases}
g_{2D}+\delta g \  \alpha t, &{ 0<\alpha t\leq 1}\\
g_{2D}+\delta g , &{\alpha t>1},
\end{cases}
\end{equation}
where $\delta g$ is the difference and $\alpha$ is the rate at which $g$ is varied over a time $\tau=1/\alpha$. In the presence of the Gaussian barrier, $g(t)$ leads to the formation of DSs in the condensate [see Fig.\ref{fig:nd}(a)]. The properties of the soliton depend on $V_0$, $d$, $\alpha$ and $\delta g$. Hence, by controlling these parameters, the DSs can be generated in a more controlled way. As we have found, both increment and decrement in $g$ lead to the formation of solitons. If the ramping of $g(t)$ fulfills the adiabatic criteria \cite{adia_SR_int_07,adia_SR_pra_02}, the condensate adiabatically follows the instantaneous ground state, and as a result, no solitons are generated. This has two immediate consequences on the properties of DSs; as $\alpha$ decreases, for a fixed $\delta g$: (i) the number of solitons decreases  and (ii) they get shallower and shallower [see Fig.\ref{fig:nd}(b)]. On a similar note, for a fixed $\alpha$, the soliton depth  and also the number of solitons increases with increase in $\delta g$.

Note that, even for sufficiently short ramping time $\tau$ with $|\delta  g|=g/2$, the solitons are very shallow [see Fig.\ref{fig:nd}(b)]. That means, the formed solitons travel close to the speed of sound, that may cause difficult to track them in the experiments. In the particular example shown in Fig.\ref{fig:nd} with $\tau=0.4$ ms, the life time of the solitons is about $15-20$ ms. In general, the life time depends on the quenching time as well as the trap geometry. Below, we consider the periodic modulation of interactions, there the soliton depths are augmented significantly, with qualitatively new features emerging.

\subsubsection{$g$ periodic in time}

\begin{figure}[hbt]
\centering
\includegraphics[width= 1.\columnwidth]{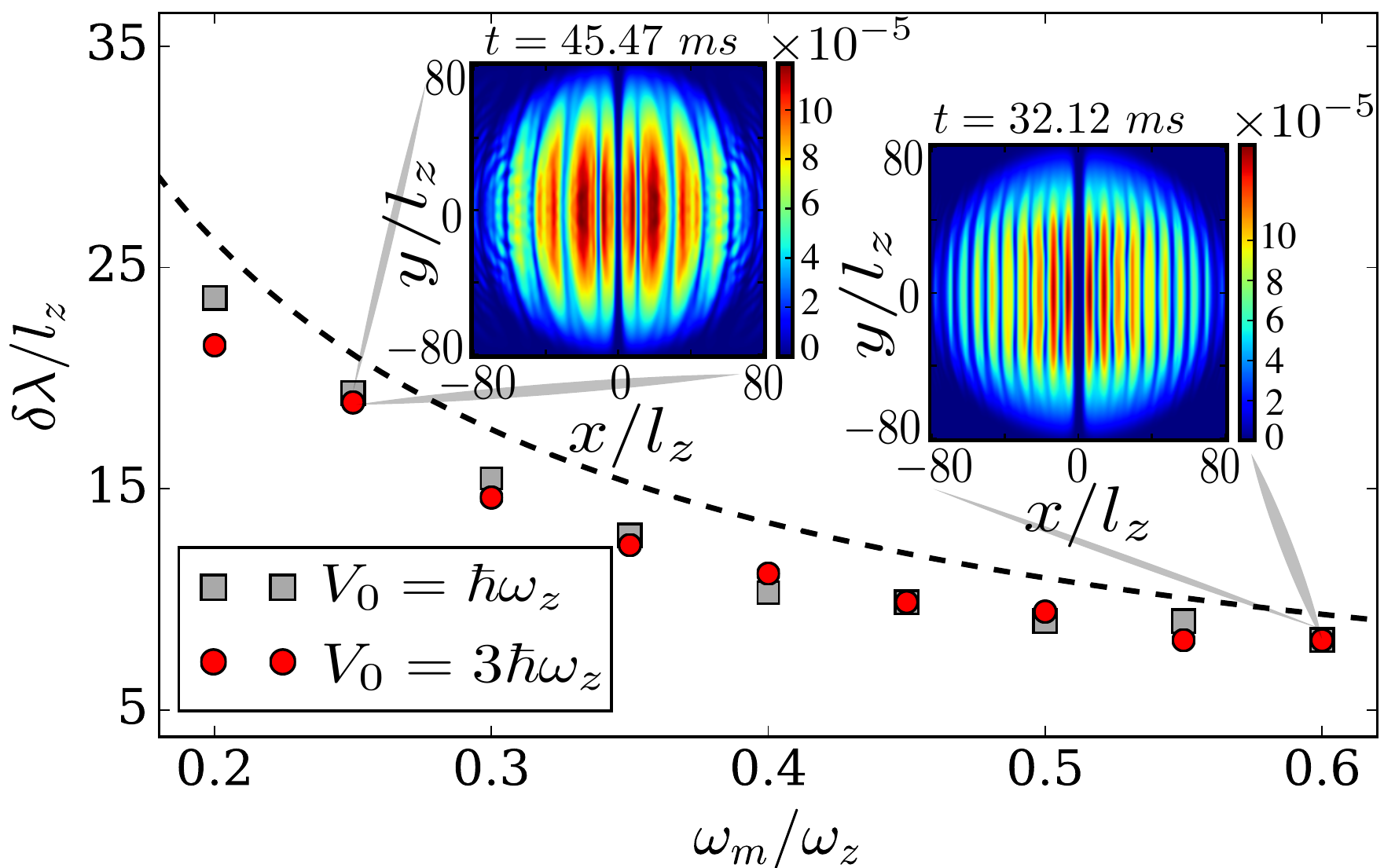}
\caption{\small{(Color online) DS lattice periodicity vs modulation frequency, $\omega_m$ for $V_0=1\ \hbar\omega_z$ (squares) and for $V_0=3 \hbar\omega_z$ (circles) with $\gamma=0.2$ and $b=0.86 \ l_z$. The two insets show the density snapshots for  $V_0=3 \ \hbar\omega_z$ at $\omega_m=0.25 \ \omega_z$ and $\omega_m=0.6 \ \omega_z$. The dashed line shows, $1/k=1/\epsilon^{-1}(\omega_m)$, the wavenumber corresponding to the Bogoliubov frequency $\omega_m$. The BEC setup is same as that of the Fig. \ref{fig:d1}.}}
\label{fig:ld} 
\end{figure}

In this section, we take $g_{2D}(t)=g_{2D}[1+2\gamma\sin(2\omega_m t)]$, where $\gamma$ and $\omega_m$ are respectively the amplitude and frequency of the modulation. As we show below, the numerical results for the dynamics of the system depend crucially on whether the initial state has a noise or not. In experimental setups always, there exist noise in the form of thermal or quantum fluctuations. They play an important role in, for instance, spontaneous symmetry breaking (SSB). A prime example for SSB is the formation of FPs in condensates by the parametric modulation of interactions \cite{farad_prl_02, farad_expt_07}, in which the noise triggers the population of Bogoliubov modes according to the wave number selection or parametric resonance condition: $\epsilon(k)=\hbar\omega_m$. The Bogoliubov spectrum of a 2D homogeneous condensate is $\epsilon(k)=\sqrt{E_k\left(E_k+2g_{2D}n_0\right)}$, where $E_k=\hbar^2k^2/2m$ with $k$ being the quasi-momentum. For sufficiently small values of $V_0$ and large values of $d$ we can still fairly approximate the spectrum of the trapped case with that of the homogeneous case  within LDA for $k>2\pi/R_{x,y}$. This low momenta cutoff introduces a low-frequency limit for the modulation frequency to observe pattern formation. Also, it has to be less than $\omega_z$ to preserve the 2D nature of the condensate.

\begin{figure*}[hbt]
\centering
\includegraphics[width= .9\textwidth]{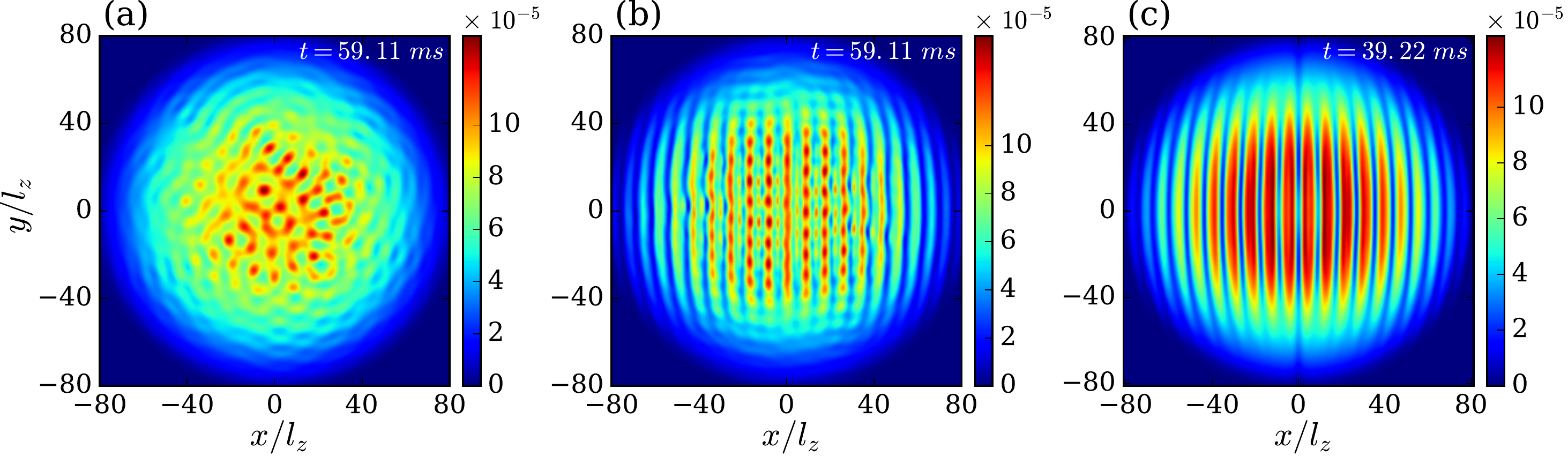}
\caption{\small{(Color online) (a) Faraday patterns for $V_0=0$. (b) Co-existence of Faraday patterns and 2D DS lattice for  $V_0=0.003\hbar\omega_z$. (c) Formation of pattern-free, transient DS lattice for $V_0=0.1\hbar\omega_z$. For all plots $\alpha=0.2$ and $\omega_m=0.6\omega_z$. The BEC setup is same as that of the Fig. \ref{fig:d1}. The extended time evolution of (b) and (c) are shown in Fig. \ref{fig:dm}.}}
\label{fig:pst} 
\end{figure*}

 {\em Noise free}:- With out noise, the FPs are not observed with the periodic modulation of the $g_{2D}$. Even for small $V_0$ we observe the formation of 2D DSs. In contrary to the linear quench, for the periodic case the depth of the gray soliton increases as a function of time and also the DSs are created continuously from the centre of the condensate. As a result, a transient 1D lattice of 2D DSs is formed after sufficiently long time, with the lattice periodicity depends crucially on the modulation frequency $\omega_m$.  Later, the lattice melts due to the SI and results in a denser gas of vortices compared to the previous cases. The periodic modulation eventually results in heating and the destruction of the condensate. Hence the formed vortices may decay thermally as well. The rate of heating depends on both the amplitude and the frequency of modulation.
  
 Note that, due to the inhomogeneity in the  density, the periodicity at the centre is different from that of the edges of the condensate. The lattice periodicity, provided by the wavelength $\delta\lambda$, calculated at the centre, as a function of $\omega_m$ for different $V_0$ is shown in Fig. \ref{fig:ld}. Interestingly, the behaviour of $\delta\lambda$, is found to be very close to the wave number corresponding to the  Bogoliubov frequency $\omega_m$ (dashed lines in Fig. \ref{fig:ld}) of the condensate at $V_0=0$. This indicates that, identical to the case of FPs, there is a wave number selection for DS lattice, closely connected to the Bogoliubov spectrum of the unmodulated condensate. This is also evident from Fig. \ref{fig:ld} that $\delta\lambda$ hardly changes with $V_0$ for sufficiently large $V_0$.
 
\begin{figure}[hbt]
\centering
\includegraphics[width= .9\columnwidth]{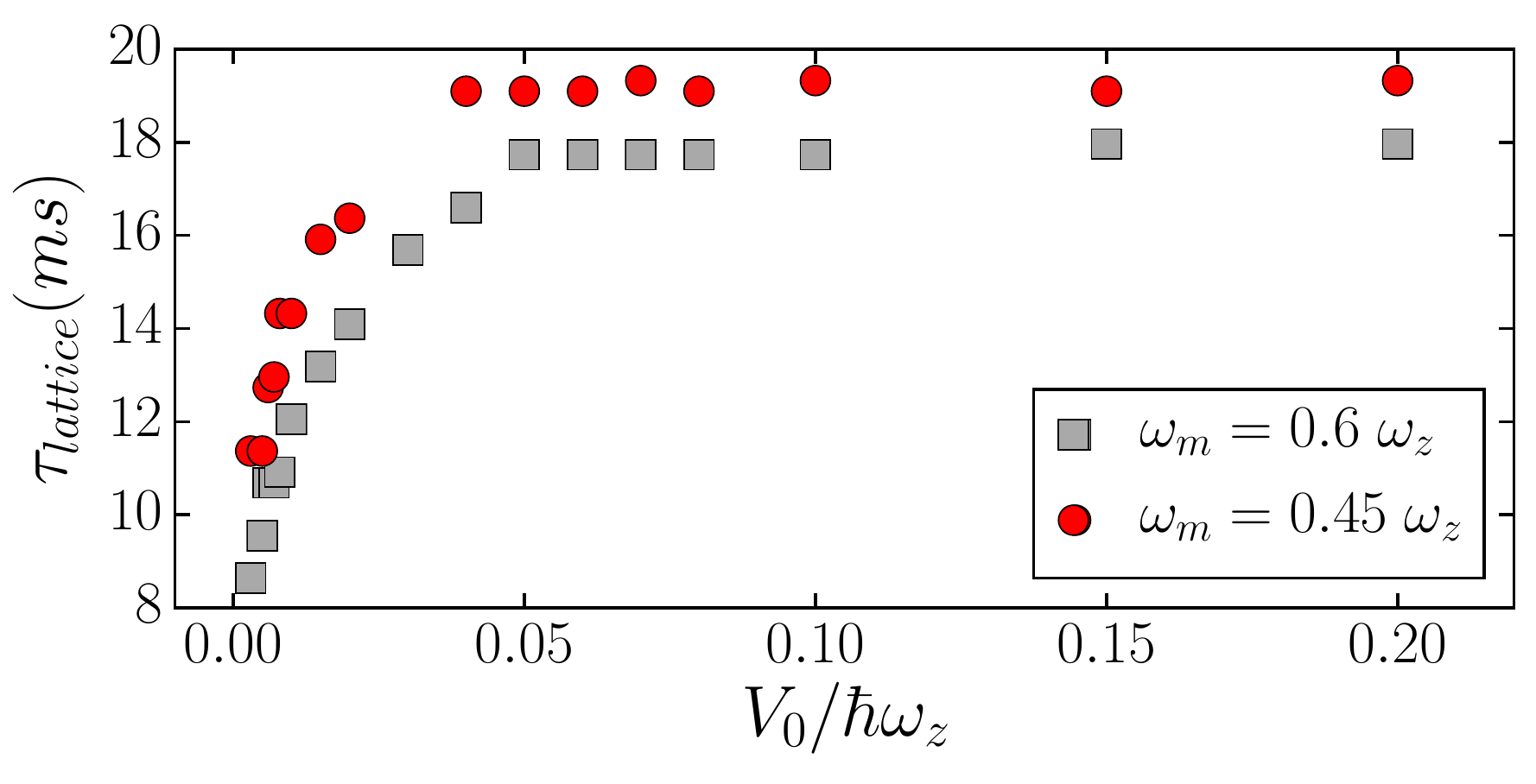}
\caption{\small{(Color online) The life time of the lattice ($\tau_{lattice}$) as a function of $V_0$  for two different $\omega_m$ for the condensate confined with frequencies: $\omega_x=\omega_y=2\pi\times 10$ Hz and $\omega_z=2\pi\times 700$ Hz.  The lifetime $\tau=\tau_2-\tau_1$, where $\tau_1$ is the time at which the lattice is stabilized to a time-independent periodicity, and $\tau_2$ is the time at which the outer DSs are unstable.The setup is same as that for the Fig. \ref{fig:pst}.}}
\label{fig:lt} 
\end{figure}

{\em With noise}:- The presence of the noise leads to an interesting scenario where the soliton formation and FPs co-exist, depending on the strength of $V_0$. In the absence of the potential barrier ($V0=0$) the FPs emerge as a result of parametric modulation of $g_{2D}$ [see Fig. \ref{fig:pst}(a)], with the wave number selection governed by the Bogoliubov modes as discussed above. For small and non-zero $V_0$, the pattern and DS lattice co-exist as shown in Fig. \ref{fig:pst}(b). Due to the presence of DSs, the FP exhibits a 1D character with density modulations along the $y$ axis, and have the same periodicity as that of the DS lattice. Note that, the presence of pattern reduces the {\em darkness} of the DSs. For large $V_0$ we see the formation of a pattern-free DS lattice [see Fig. \ref{fig:pst}(c)], which eventually undergoes snake instability and breaks up into a gas of VDs. With large $V_0$, the soliton formation is enhanced, resulting in the fast formation of the lattice, which suppresses the pattern formation in the condensate. The 
 life time of the lattice ($\tau_{lattice}$) as a function of $V_0$ is shown in Fig. \ref{fig:lt} for two different modulation frequencies. At small values of barrier height, $\tau_{lattice}$ depends strongly on $V_0$, and it saturates for sufficiently large values of $V_0$. 

\begin{figure*}[hbt]
\centering
\includegraphics[width= 1.\textwidth]{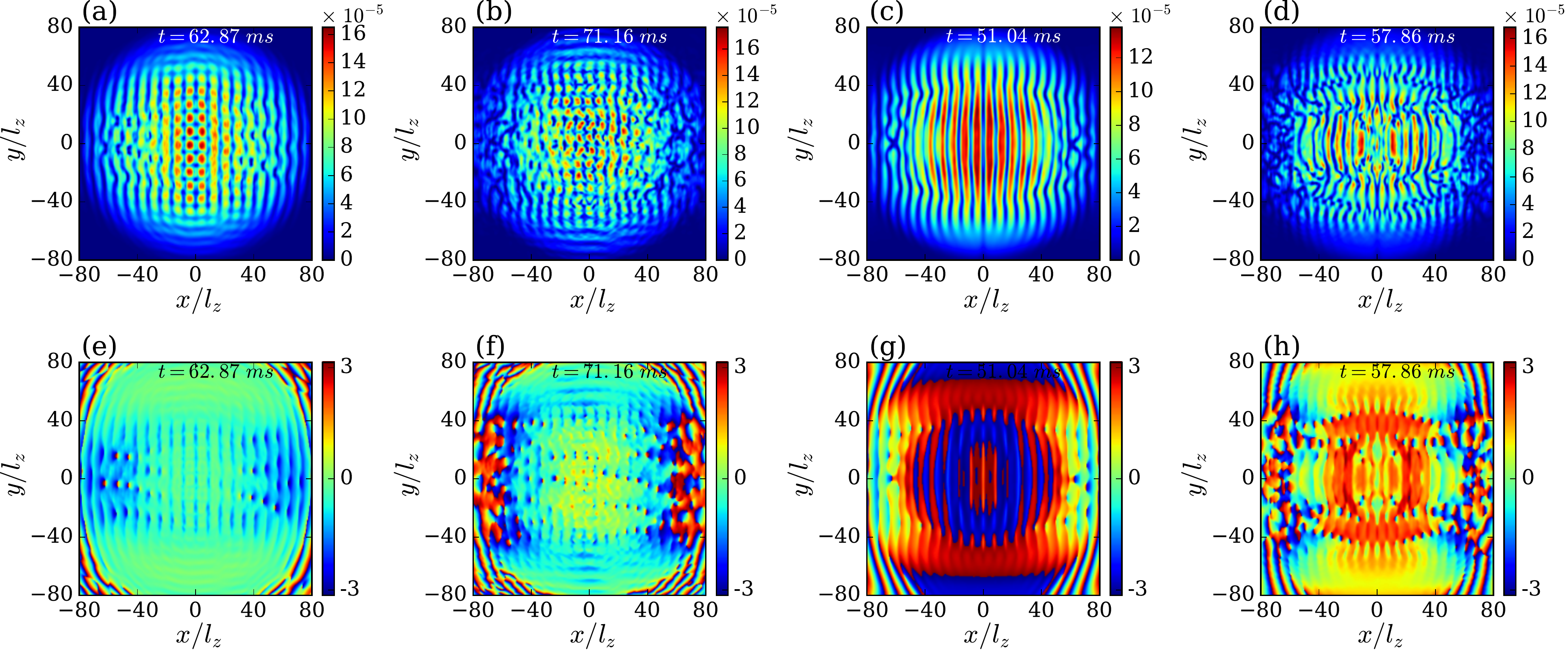}
\caption{(Color online) \small{(a) and (b) show the extended time evolution of the state shown in Fig. \ref{fig:pst}(b). The phase plots corresponding to (a) and (b) are shown in (e) and (f) respectively. (c) and (d) show the extended time evolution of the state shown in Fig. \ref{fig:pst}(c), with corresponding phase plots in (g) and (h). (a) and (b) are for $V_0=0.003\hbar\omega_z$ whereas (c) and (d) are for $V_0=0.1\hbar\omega_z$. All other details of the setup are same as that of Fig. \ref{fig:pst}.}}
\label{fig:dm} 
\end{figure*}

The instability dynamics crucially depends on the strength of $V_0$. For any $V_0$, the instability emerges first at the solitons near the boundary, since they are created before those at the centre of the condensate. Interestingly, for low $V_0$ the presence of Faraday patterns give rise to a characteristically different instability dynamics for the DSs at the centre of the condensate. The DS stripes embedded with FP break up into vortexoniums through a transient square pattern of density peaks [see Fig.\ref{fig:dm}(a)-(b)]. In contrary, for large $V_0$ when there is no FP, they directly break up into VDs  [see Fig.\ref{fig:dm}(c)-(d)]. The corresponding phase plots are also shown in Fig.\ref{fig:dm}(e)-(h).

Finally, we want to comment that similar results can be obtained by making either the potential barrier or the trapping frequency become  time dependent instead of interactions.

 \section{Conclusion}
 \label{con}
 In conclusion, we numerically analyzed the long-time dynamics of SI of 2D DSs in a TF condensate of rubidium atoms, with contact interactions. The DS is dynamically generated in the BEC by an initially imprinted phase gradient. The soliton properties as well as the nature of SI and post-instability dynamics depend substantially on the spatial width of the phase gradient. The VDs emerging from the unstable DS exhibits interesting dynamics such as the formation of vortexoniums, vortex annihilation, and exchange processes. The effective decay in the number of vortices is indirectly influenced by the spatial extension of the phase gradient. Alternative techniques for the DS generation, based on time-dependent interactions and an external Gaussian barrier are discussed. For the linear variation of interactions, the properties of the DS can be controlled by quench time. A transient DS lattice emerged in the condensate for the parametric modulation of interactions, with lattice periodicity depending on the frequency of modulation. Interestingly, the scenario is richer by the co-existence of Faraday patterns and the transient DS lattice together. The SI dynamics of the soliton-lattice is significantly modified if FP is present. We also provide other techniques to generate DSs, which would provide similar results.
 
 \section{Acknowledgments}
 UDR would like to thank the Department of Science and Technology, Govt. of India 
for funding through grants from EMR/2014/000365 and Nano Mission. R.N. acknowledges funding by the Indo-French Centre for the Promotion of Advanced Research.
  \bibliographystyle{apsrev}
\bibliography{liball.bib}
\end{document}